

\magnification=1200

\advance\hoffset by -3mm       
\advance\voffset by -3mm       
\parindent 10pt


\font\titlef=cmbx12 scaled\magstephalf
\font\sectf=cmbx10
\font\subsectf=cmsl10
\font\authorf=cmbx10
\font\affont=cmsl10


\def\title#1#2{{\centerline{\titlef{#1}}}{\centerline{\titlef{#2}}}\vskip
.5cm}
\def\author#1{{\centerline{\authorf{#1}}}\medskip}
\def\affiliation#1#2#3{{\centerline{\affont{#1}}}{\centerline{\affont{#2}}}

{\centerline{\affont{#3}}}\vskip .5cm}
\def\abstract#1{{\bigskip\centerline{\bf
ABSTRACT}}\medskip{#1}\vfill\eject}

\newcount\sectionnumber
\newcount\eqnum

\newcount\subsectionnumber

\def\section#1{\global\subsectionnumber=0\global\eqnum=0
\medskip\global\advance\sectionnumber by 1
${\hbox{\bf{\the\sectionnumber.}}}\,\,{{\hbox{\sectf{#1}}}}$\medskip}
\def\subsection#1{\medskip \global\advance\subsectionnumber by 1
${\hbox{\sl{\the\sectionnumber-\the\subsectionnumber}}}\,\,
{\underline{\hbox{\subsectf{#1}}}}$\medskip}


\def\num{(\the\sectionnumber.\the\eqnum)}


\def\eqnu#1{\global\advance\eqnum by 1\eqno
{(\the\sectionnumber.\the\eqnum)}}

\def\aeqnu#1{\global\advance\aeqnum by 1
\eqno{\scriptstyle{\mit(A\the\appnumber.\the\aeqnum)}}}




\def\label{\edef}


\def\e{{\rm{e}}}
\def\im{{\rm{i}}}
\def\M{{\cal{M}}}
\def\B{{\cal{B}}}
\def\Z{Z\!\!\!Z}
\def\R{I\!\!R}
\def\atbound{{\big\vert_\B}}


%



\hfill{QMUL-PH-02-03}

\hfill{SWAT-TH/03-371}

\hfill{hep-th/0303229}

\bigskip

\title{Charges and fluxes in Maxwell theory}{on compact manifolds
with boundary}

\author{Marcos Alvarez}
\affiliation{Physics Department, Queen Mary, University of London}
{Mile End Road, E1 4NS London, UK}{\tt m.alvarez@qmul.ac.uk}
\author{David I. Olive\footnote*{\tenrm Presently at
NORDITA, Blegdamsvej 17, DK2100 Copenhagen}}
\affiliation{Physics Department, University of Wales Swansea}
{Singleton Park, Swansea SA2 8PP, UK}{\tt d.olive@swansea.ac.uk}
\abstract{We investigate the charges and fluxes that can occur in
higher-order Abelian gauge theories defined on compact space-time 
manifolds with
boundary. The boundary is necessary to supply
a destination to the electric lines of force emanating from brane sources,
thus allowing non-zero net electric charges, but it also introduces new
types of electric and magnetic flux. The resulting structure of currents,
charges, and fluxes is studied and expressed in the language of
relative homology and de Rham cohomology and the corresponding abelian
groups. These can be organised in terms of a pair of
exact sequences related by the Poincar\'e-Lefschetz isomorphism and by
a weaker flip symmetry exchanging the ends of the sequences.
It is shown how all this structure is brought into play
by the imposition of the appropriately generalised Maxwell's equations.
The requirement that these equations be integrable
restricts the world-volume of a permitted brane (assumed closed)
to be homologous to a cycle on the boundary of space-time. All electric
charges and magnetic fluxes are quantised and satisfy the Dirac
quantisation
condition. But through some boundary cycles there may be
unquantised electric fluxes associated with
quantised magnetic fluxes and so dyonic in nature.}

\section{Introduction}

In the search for a unified theory of particle interactions encompassing
both the standard model and Einstein's theory of gravity the most promising
candidate seems to be the superstring and M-theories which require
space-time
to have dimensions $10$ and $11$ respectively for internal consistency.

A common feature of these is the presence of states known
as \lq\lq $p$-branes'', objects which, classically at least,
can be pictured as extended objects resembling $p$-dimensional
surfaces (or volumes) in space. As time evolves these sweep out
surfaces (or volumes) of one dimension higher, $p+1$.

When $p=0$, the object is simply a point particle tracing out a
world-line, $w$, in space-time. It has a geometrically natural
interaction with Maxwell's electromagnetic field specified by
the addition of a term in the action taking the schematic form
$$
\lq\lq q\int_w A\hbox{''}.\eqnu{inter}
$$\label\inter{\num}\par\noindent
The same can be done for any positive value of $p$ less than $m$,
the dimension of space-time, with the proviso that the gauge
potential $A$  now has degree $(p+1)$, matching the fact that
the world-volume, $w$, has $(p+1)$-dimensions. When $p$ equals one,
so that the brane is a string, $A$ is the familiar Kalb-Ramond [1974] gauge
potential.

Naively Stokes' theorem implies that expression \inter\ is unchanged
when $A$ is altered according to
$$A\rightarrow A+d\chi, \eqnu{gauge}$$\label\gauge{\num}\par\noindent
where $\chi$ is of degree  $p$ and arbitrary, if it assumed that $w$ is
closed, i.e. a $(p+1)$-cycle. This generalised gauge invariance
suggests that an important physical role would be played by the
following quantity which is invariant with respect to \gauge:
$$
F=dA. 
\eqnu{fs}
$$\label\fs{\num}\par\noindent
This is the $(p+2)$-form field strength, reducing to the familiar
one of Maxwell when $p=0$.

The most natural equations of motion for $F$ take the form:
$$dF=0,\eqnu{maxa}$$\label\maxa{\num}\par\noindent
$$d*(hF)=*j, \eqnu{maxb}$$\label\maxb{\num}\par\noindent
in exterior calculus notation, although there are more elaborate
possibilities. In order to include a common feature
of supergravity/superstring theories we have admitted
the presence of a positive scalar function of
scalar fields, $h(\phi)$,
in \maxb, that equals unity in vacuo. Apart from this feature,
these are Maxwell's equations generalised in the
form envisaged by Hodge, and $*$ denotes
his duality operation, constructed by means of a metric on space-time,
here assumed
to be a fixed background [Hodge 1952, Flanders 1963]. We shall
henceforth refer to them as Maxwell's equations.

The inhomogeneous Maxwell equation \maxb\ will play an important role
in what follows irrespective of the detailed form of the quantity $h$
as long as it reduces to unity in vacuo and we shall not have to consider
equations of motion for the scalar fields. The quantity $j$ is the
\lq\lq  electric current'' and is a $(p+1)$-form, so possessing the
same degree as $A$. By virtue of \maxb\ and the nilpotency of the
exterior derivative, $d$, it has to be \lq\lq conserved'' so that
$$
d^*j=0,\eqnu{cc}
$$\label\cc{\num}\par\noindent
and we shall always suppose this.

Such field theories are indeed part of modern superstring/M-theory
and it is therefore important to understand their properties by
answering the questions listed below, particularly when the
background space-time is taken to be topologically complicated.
But these equations are only a part of the larger theory and
not the whole. In this subtheory,  no account need be taken of
supersymmetry
and the values of $p$ and $m$, the dimension of space-time $\M$,
can treated as arbitrary. Many special features of these subtheories
have become familiar, [Nepomechie 1985, Teitelboim 1986, Henneaux and
Teitelboim 1986, Deser, Gomberoff, Henneaux and Teitelboim 1997] but
our aim is to uncover yet more general structure as will be seen.

When $p$ vanishes and $m$ equals $4$ there are three notions familiar
since the times of Faraday, electric charge, electric flux and
magnetic flux. (Magnetic charge is excluded by \maxa\ since magnetic
current is). These three quantities are all conserved, that is unchanged
by various sorts of evolution, including that in time. It will be seen
to be  important to distinguish the notions before examining the
possibility of any relations between them.

Allowing $p$ and $m$ to be arbitrary,
the physical questions considered in this paper concern:
\vskip5pt
\noindent (1) the classification and enumeration
of the independent charges and fluxes

\noindent (2) the understanding of how the Maxwell equations \maxa\ and
\maxb\
relate the notions of electric flux and charge,

\noindent (3) the determination of the possible numerical
values of these charges and fluxes,

\noindent (4) the understanding of how quantum theory can
relate the values of electric and magnetic conserved quantities
(yielding the generalised Dirac quantisation condition).
\vskip5pt
The answers turn out to be more subtle and interesting than
we had expected and it is this that motivates this presentation.
It is found important to resist the  common temptation to simplify by
taking the space-time manifold, $\M$, to be closed as this results in
oversimplification. When account is taken of the generalised Maxwell
equations, \maxa\ and \maxb, all electric charges and fluxes then vanish,
leaving magnetic fluxes as the only available conserved quantities, as
Henneaux and Teitelboim [1986] have pointed out.

Thus it is essential to allow space-time, $\M$, to possess a boundary,
$\B$, a manifold of dimension one less, interpreted as corresponding
to the \lq\lq points at spatial infinity'' through which
\lq\lq electric field lines'' may escape, thereby furnishing
a potentially non-trivial flux. When this is done, the answers
to the physical questions above are provided by a set of results
in pure mathematics whose physical relevance is, we believe,
hitherto unappreciated by physicists. Once we have established
the appropriate definitions we find the connections, made more
precise in the text:
\vskip5pt
electric charges $\Leftrightarrow$ relative homology of space-time,

electric fluxes $\Leftrightarrow$ absolute homology of the boundary
of space-time,

magnetic fluxes $\Leftrightarrow$ absolute homology of space time.
\vskip5pt
All these charges and fluxes are expressed as integrals over
some sort of cycle in space-time and homology deals with the
classification of these cycles in the way that is appropriate
to the physics. There are precisely three types of homology
in the situation just described and all three play a physical role
according to the connections just listed. Moreover the
relationships between the different sorts of conserved quantity
correspond to relationships between these different sorts of homology.

All of the aforementioned types of homology class form elements of an
abelian group, the appropriate homology group, $H_*$, say.
Taking into account
all  values of $p$ that are possible in the  given  fixed
background space-time, $\M$, these abelian groups can be
arranged in a certain order such that there is
a natural homomorphism acting between
successive members. This provides a sequence with the property
of being exact, that is, at each stage, the homology group,
$H_*$, possesses a subgroup, $K_*$, say, that is at the same
time the kernel of the succeeding homomorphism and the image
of the preceding one. This is the exact sequence of relative
homology (of space-time). A more refined classification of
the physical notions of charge and flux will depend on the
distinction between the subgroup
$K_*$ and the coset group $H_*/K_*$ within each homology group $H_*$.
This structure is explained in more detail in the text as
it becomes relevant to the development of the physical arguments and
particularly in Sections 7 and 10, as well as the Appendix. Relevant
mathematical background together with more detail
can be found in [Schwarz 1994 and Massey 1991].

Each homology group, $H_*$, is abelian, and usually of infinite order.
For reasons explained they are essentially discrete lattices of finite
dimension, $b_*$ which is known as  the Betti number.
The number of linearly independent charges and fluxes
will be expressible in terms of Betti numbers in a surprisingly
complicated way that we shall determine. These results will answer
the first two of the physical questions listed above.

An important subtlety is that although the definition of the
conserved electric charge as an integral over the conserved current,
$j$ works irrespective of whether or not the generalised Maxwell
equations \maxa\ and \maxb\ are assumed to hold, the counting of the
charges
does depend on this choice, being more complicated when they
do hold, as they should when account is taken of physical relevance.
For example,
when space-time is closed, all possible non-trivial electric
charges are forced to vanish by the equations \maxa\ and \maxb.
The point is that there exist conserved currents on space-time
for which it is impossible to integrate \maxb\ to obtain a field strength,
$F$. Consequently these currents will be forbidden on the physical
grounds that the field strengths must exist.
It is the aforementioned exact sequence of relative
homology that clarifies the occurrence of this  phenomenon
as explained in section 4 and amplified later.

Answering the third of the physical questions listed above
requires an explicit form of the conserved current, $j(w)$,
due to a $p$-brane with world-volume $w$ as implied by \inter\ and
\maxb\ together. This is provided by a singular differential
form involving Dirac $\delta$-functions whose support is $w$.
Then the electric charge associated with integrating over a
relative cycle $S$ is $q$ times the intersection number of
$S$ and the absolute cycle, $w$. The coefficient $q$ is
defined by \inter\ and the intersection number is well defined
as $w$ and $S$ have dimensions summing to $m$, that of space-time.
As this intersection number is unchanged by homologies of both
$w$ and $S$, it is defined on their homology classes.
Since the groups formed by these classes are essentially
lattices whose dimension is the relevant Betti number,
it follows that the intersection data is encoded in the
intersection matrix, $I$, formed of the intersection
numbers between elements of bases of the two lattices.
This matrix, $I$, has integer entries, is square and unimodular (that is,
has determinant equal to $\pm1$), the latter two properties
being consequences of \lq\lq Poincar\'e-Lefschetz duality'',
another feature of the exact sequence of relative homology.

So far this analysis does not use the \lq\lq Maxwell equations'',
\maxa\ and \maxb,
and hence applies whether or not they are chosen to hold. If not,
the electric charges take values equal to an integer times $q$.
Conversely the unimodularity of the intersection matrix means that
there exist brane configurations realising all possible values of
these sets of values.

If Maxwell's equations are chosen to hold, as they should, the situation
is more complicated as many potential electric charges are forced to
vanish,
apparently contradicting the unimodular property of the intersection
matrix.
The resolution of this paradox depends on the recognition that some brane
configurations are forbidden as they yield conserved electric currents
for which the Maxwell's equations \maxa\ and \maxb\ cannot be
integrated to yield a
field strength. This is explained in more detail in section 8 and
requires the intersection matrix to have a more detailed structure
than so far apparent. This is revealed by writing it in block form
according to the kernel subgroup, $K_*$, of each $H_*$, and the
coset $H_*/K_*$. One block has to vanish identically and this is
verified explicitly in section 10 and the Appendix. This
leaves square matrices on the block diagonal each
of which  have to be unimodular.

The upshot is that the only brane configurations that are allowed
by the integrability requirement are those that are homologous to
cycles in the boundary, $\B$, of space-time, $\M$. The surviving
electric charges again take values that are integral multiples of
$q$. Conversely there are allowed brane configurations that realise
all possible sets of these values.

Another phenomenon quantified by the exact sequence of homology
is the existence of electric fluxes which are not equal
to electric charges and hence not quantised. Through the same cycles
there may flow quantised magnetic fluxes of the field coupling to the 
brane
dual to that coupling to the electric field so that the overall effect is
suggestive of something  dyonic.

All results so far are \lq\lq classical'', invoking no quantum theory.
Taking the latter into account requires that the schematic term \inter\
in the action be unambiguous when suitably exponentiated.
This constrains the values of the magnetic fluxes to satisfy
a generalisation of  Dirac's celebrated quantisation when
compared with any of the electric charges [Dirac 1931, Wu and Yang 1975,
Alvarez and Olive 2000].

The resulting picture is beautifully consistent yet unexpectly rich.
Nevertheless our analysis made a number of implicit simplifications
compared with the full superstring theory that are so far unavoidable.
Some of these are listed in the conclusion, section 11,
and it is hoped that a subsequent
elaboration of our present methods
will lead to answers removing these assumptions.

A technical Appendix extends the idea of a distribution
valued form associated with a bulk cycle (such as the brane world-volume)
to chains both in the bulk and
on the boundary. These constructions are used to derive the weak form
of Poincar\'e-Lefshetz duality used in establishing the vanishing of
an off-diagonal block of the previously mentioned intersection matrix.

Relative topology has been used previously to discuss certain aspects
of branes in M-theory. A partial description of the role of relative
cohomology in the classification of charges in generalised Maxwell
theory was sketched in section 2 of [Moore and Witten 2000]. In
[Kalkkinen and Stelle 2003] relative cohomology is used to present a
geometric description of certain brane intersections in M-theory. A
similar analysis of D2-branes in Wess-Zumino-Witten theory can be
found in [Figueroa-O'Farrill and Stanciu 2001].

\section{First notions}

Taken as given is a fixed background space-time $\M$,
assumed oriented and compact, but possibly
of  complicated topology. It has dimension $m$ and
initially it is assumed to be closed.  On it is defined a field strength
$F$
that is a  $(p+2)$-form satisfing the generalised Maxwell equations \maxa\
and \maxb. According to the first of these $F$ is closed so that
locally there is defined a
$(p+1)$-form gauge potential $A$, \fs, with a gauge ambiguity with
respect to the gauge
transformations \gauge\  where $\chi$ is a $p$-form also defined locally.
The quantity $j$ is a $p+1$-form denoting
the electric current due to the matter degrees of freedom. For the time
being
it does not have to be assumed that it has the form that \inter\ would
imply.

Electric current
conservation is the statement that $^*j$ is a closed form on $\M$, \cc.
This follows from  the above Maxwell equation  \maxb\ as
$d^2$  vanishes, but we shall assume its validity even when Maxwell's
equations are disregarded.

The first notion of an electric charge is associated with the current $j$
without any reference to the field strength, $F$. Hence Maxwell's
equations can be temporarily disregarded.
It is formulated by considering  an oriented   region $S$ that is a
$(m-p-1)$-chain over which it is possible to integrate the matching
form $*j$:
$$
Q(S)=\int_S*j
\eqnu{charge}
$$\label\charge{\num}\par\noindent
Conventionally the region $S$ would be thought of as \lq\lq
space-like'' but this is not essential.
The virtue of the definition \charge\ is that it is insensitive to
alterations of $S$ by
homologies that preserve its boundary, $\partial S$. Thus, if
$S'=S+\partial C$, so $\partial S'=\partial S$,
$$
Q(S')=Q(S)
$$
as $Q(\partial C)=\int_{\partial C}*j=\int_{C}d*j=0$,
using Stokes' theorem and current conservation \cc. This establishes a
good sense in which the charge
$Q$ is conserved. The disadvantage of this definition is that the
regions $S$ for which the charge is defined lack
any real homological significance unless it is assumed that $S$ is
closed, $\partial S=0$. Now the result means that each electric
charge, $Q(S)$, is preserved by homologies of $S$, that is,
unchanged by the kinds of evolution associated with these homologies.
Homologous surfaces form absolute homology classes which
themselves form an abelian
group under addition of surfaces, in this case the absolute homology group
of $\M$, denoted
$H_{m-p-1}(\M;\Z)$. Without any field strengths satisfying the
Maxwell equations this would be the end of the story as there would be
no fluxes to consider.

Since field strengths are included, it is necessary to consider the effect
of applying Maxwell's equation \maxb:
$$
Q(S)=\int_S*j=\int_S\,d*(hF)=\int_{\partial S}*(hF)=0
$$
as $\partial S$ vanishes. Thus all electric charges vanish when
Maxwell's equations hold on a closed space-time, $\M$.
In physical terms, the problem is that the Maxwell equation \maxb\
attaches electric
lines of force to the electric charge distribution and these lines
have nowhere to go. Mathematically the point is that
when the conserved electric current, $j$, is such that any $Q(S)$
fails to vanish,
it is impossible to integrate \maxb\ to obtain the field strength $F$ on
$\M$.
This is unaceptable on physical grounds.

An obvious remedy is to provide a destination for the lines of
force by allowing space-time, $\M$, to be non compact,
and this will be considered next. But it will remain necessary
to check the integrability of Maxwell's equations in the sense just
described,

\section{Electric Charges and Relative Homology}

Instead of allowing space-time, $\M$, to be non-compact, as just suggested,
we shall do something slightly different and keep it compact but allow
it to have a non-trivial boundary, $\B=\partial\M$, of
one dimension less. This can be thought of as comprising those points
at spatial infinity through which electric lines of force may escape.
On the other hand, electric current,
$j$, is not allowed to escape, that is its Hodge dual, $*j$, is assumed
to be localised and this is expressed by the boundary condition:
$$
* j\atbound=0.
\eqnu{twob}
$$\label\twob{\num}\par\noindent
More precisely this means that the restriction of the differential form
$*j$ to
$\cal B$ vanishes. Thus the normal components of $j$ vanish on $\B$.
In addition, it is assumed that the scalar function, $h$, occurring in
\maxb,
takes its vacuum value on $\B$:
$$
h\atbound=1\eqnu{hb}
$$\label\hb{\num}\par\noindent
Of course Maxwell's equations, \maxa\ and \maxb\ and also current
conservation,
\cc\ still hold on $\M$, or as we shall say, in the bulk.

The same expression \charge\ for the electric charge holds good except that
now, instead of assuming $\partial S$ vanishes, we assume that it lies in
$\B$, and so $S$ has become  what is called a relative cycle.
Suppose that $S$ is altered by a relative homology:
$$
S\rightarrow S'=S+\partial C+\beta, \qquad
C\in \M, \qquad \beta \in \B.
$$
Then $Q(\partial C)=\int_{\partial C}*j=\int_C d*j=0$, by
Stokes' theorem and \cc, while $Q(\beta)=\int_{\beta}*j=0$ by \twob. So
$$
Q(S)=Q(S')\qquad \hbox{if } S\sim S'
\eqnu{four}
$$\label\four{\num}\par\noindent
in relative homology $\M$ mod $\B$. In particular $Q(S)$ vanishes
if $S\sim 0$. Thus the electric charge is well defined  as an integral
over relative homology classes, denoted $[S]$ and forming an abelian
group, $H_{m-p-1}(\M,\B;\Z)$. This is one sense in which the electric
charges are conserved. The abelian group structure arises because two
like relative cycles can be added to form a third. According to
\charge\ this addition law is respected by the electric charges:
$$
Q([S])+Q([S'])=Q([S+S'])=Q([S]+[S'])
$$
and this furnishes another sense in which they are conserved.

Some elements of this homology group have finite order and are called
torsion elements. Thus if $S$ is not trivial, that is not relatively
homologous to $0$, yet has the property that there exists an  integer
$n$ such that $n[S]=[nS]$ is trivial, then, by the above,
$Q([S])=Q([nS])/n=0$. Altogether such
elements form a finite abelian subgroup $T$, (the torsion group),
which can be divided out
of $H_{m-p-1}(\M,\B;\Z)$ to form a free group
$$
F_{m-p-1}(\M,\B;\Z)=H_{m-p-1}(\M,\B;\Z)/T
$$
which can be regarded as a lattice of finite dimension
$b_{m-p-1}(\M,\B)$. This dimension is the corresponding Betti number.
Because
there are no contributions from torsion
elements,  electric charges are only defined on $F_{m-p-1}(\M,\B;\Z)$.
Hence the integer $b_{m-p-1}(\M,\B)$ counts what appears to be the
number of linearly independent electric charges that can be defined on
the space-time $\M$. This conclusion is an overestimate for reasons to be
explained in the next section.

{}From now on, the conventions of this section will be adopted, and
absolute chains will be denoted by lower case Roman letters
(a,b,c ... s,t,u,v,w..), relative chains by upper case Roman letters
(A,B,C ... S,T,U,V,W..) and chains in the boundary by Greek letters
($\alpha,\beta,\gamma ... \phi,\chi,\psi..$). The letters later in the
alphabet will denote the corresponding cycles.

\section{Electric fluxes and electric charges}

The above derivation of the topological classification of electric
charges by relative homology $F_{m-p-1}(\M,\B;\Z)$,  used only the
properties \cc\ and \twob\ of the current
$j$, and not the Maxwell equations $dF=0$ and \maxb.  Current conservation
\cc\ can be regarded as a necessary local  condition for the
integrability of
the field strength $F$, given the current, $j$,  but it is not
sufficient, as
already has been  seen when space-time, $\M$, has no boundary, nor will
it be
so when it does have a boundary.

Assuming the Maxwell equation \maxb\ does hold, the electric
charge $Q(S)$ can be rewritten as an electric flux:
$$
Q(S)=\int_S*j=\int_Sd*(hF)=\int_{\partial S}*(hF)=\int_{\partial
S}*F,\eqnu{flch}
$$\label\flch{\num}\par\noindent
by Stokes' theorem and \hb. Of course $\partial S$
is in the space-time boundary, $\B$, and is a cycle
though not necessarily a boundary of a chain there, even though
it is in the bulk, $\M$.

But it is possible to provide a more general definition of electric flux
than this by considering any cycle in $\B$,  not just one that
is a boundary  of a relative cycle:
$$
\Phi_E(\phi)=\int_{\phi}*F,\qquad \phi\in\B,\qquad\partial\phi=0.
\eqnu{eflux}
$$\label\eflux{\num}\par\noindent
This extended definition works on all the absolute homology classes
of the boundary,
$H_{m-p-2}(\B;\Z)$, or, more precisely, the free parts,
$F_{m-p-2}(\B;\Z)$,
defined as before. To check, consider the absolute homology in the
boundary,
$\phi'\rightarrow\phi+\partial\gamma$, $\gamma\in\B$. Then
$\Phi_E(\partial\gamma)=\int_{\partial\gamma}*F=\int_{\gamma}d*(hF)=
\int_{\gamma}*j=0$, using Stokes' theorem and equations \maxb\ and \twob.
So indeed $\Phi_E(\phi)=\Phi_E(\phi')$ if $\phi\sim\phi'$ in
absolute homology in $\B$, in distinction to the electric charges that
appeared to correspond to relative homology.

So, since their classifications differ, electric charges and electric
fluxes must be distinguished. This distinction manifests itself in
two different physical ways.

First, not all electric fluxes are expressible as electric charges
because not all cycles on the boundary, $\B$; are boundaries of chains
on $\M$. The electric fluxes that are equal to  charges are associated with
cycles on the boundary, $\B$, that are also boundaries of chains on $\M$,
as in \flch. These classes of cycles form a subgroup of the absolute
homology group of the boundary, $\B$, $H_{m-p-2}(\B;\Z)$,  that we
shall denote
as follows:
$$
K_{m-p-2}(\B;\Z)=\{{\hbox{classes of boundary cycle $\phi$
satisfying $\phi=\partial S$ for some bulk chain $S$}}\}.
\eqnu{kb}
$$\label\kb{\num}\par\noindent

Secondly there are apparently non-trivial electric charges, $Q(S)$,
which must vanish if they are expressible as fluxes.
This happens precisely when $\partial S$ is a boundary in $\B$,
as well as in $\M$, according to Stoke's theorem applied to \flch.
The classes of these cycles form a subgroup of the relative homology
group that we shall denote as follows:
$$
K_{m-p-1}(\M,\B;\Z)=\{\hbox{classes of relative cycle, $R$,
satisfying $\partial R=\partial\alpha$, $\alpha\in \B$}\}.\eqnu{kmb}
$$\label\kmb{\num}\par\noindent
It follows that it
is the vanishing of the charges associated with these cycles that is
the extra integrability
condition on the Maxwell's equation \maxb\ in order to obtain
a field strength, $F$, given a conserved current, $j$.

To recap, electric charges defined on $K_{m-p-1}(\M,\B;\Z)$ all vanish,
leaving non-trivial charges associated with each coset element of
this subgroup. Furthermore
only those electric fluxes defined on $K_{m-p-2}(\B;\Z)$ are
expressible as electric charges. What is
happening mathematically is that the boundary operation $\partial$
mapping relative cycles to boundary cycles induces a map
$$
\partial_*:\qquad H_{m-p-1}(\M,\B;\Z) \longrightarrow 
H_{m-p-2}(\B;\Z)\eqnu{homd}
$$\label\homd{\num}\par\noindent
In fact this is  a group homomorphism with
kernel $K_{m-p-1}(\M,\B;\Z)$, \kmb, and image
$K_{m-p-2}(\B;\Z)$, \kb.  So, by Lagrange's theorem,
$$
H_{m-p-1}(\M,\B;\Z)/K_{m-p-1}(\M,\B;\Z)\equiv K_{m-p-2}(\B;\Z).
$$
and it is this group (or more precisely the free part) that
classifies the non-trivial electric charges. Applying this to
the free parts that are lattices classifying the
corresponding charges and fluxes, the number of linearly independent
electric charges is given by
$$
b_{m-p-1}(\M,\B)-s_{m-p-1}(\M,\B)=s_{m-p-2}(\B),\eqnu{chcount}
$$\label\chcount{\num}\par\noindent
explaining the overestimate mentioned previously. The integers
$s_*(X)$ are
the dimensions of the lattices specifying the free part of $K_*(X;\Z)$.

As there are $b_{m-p-2}(\B)$ linearly independent fluxes, $s_{m-p-2}(\B)$
of which are expressible as electric charges the difference,
the number $b_{m-p-2}(\B)-s_{m-p-2}(\B)$, specifies the number of linearly
independent electric fluxes that cannot be equated to electric charges of
the form \charge.

\section{Relation between the preliminary and final versions of
electric charge}
For reasons that become clear later, it is worth asking a question that
seems rather ridiculous from a physical point of view, namely how to
relate the class of electric charge obtained by integrating $*j$ over
a bulk cycle to the class
obtained by integrating over a relative cycle. The first class,
considered in our preliminary discussion still makes sense
when space-time has a boundary since a bulk cycle can
be considered as a special case of a relative cycle. The reason
the question is apparently ridiculous from a physical point of view
is that these preliminary charges do all vanish when account is
taken of Maxwell's equations as already seen.

Consider an absolute bulk $(m-p-1)$-cycle, $r$, and decompose it into
the sum of
a $(m-p-1)$-chain in $\B$ and a remainder that contains no such chain:
$$
r=R+\alpha.
$$
As $\partial r=0$, $\partial R=-\partial\alpha\in\B$,
so that $R$ is a relative cycle. Furthermore, if $r$  is trivial as
a bulk cycle, so  $r=\partial a$, then
$R=\partial a-\alpha$ and so is trivial as a relative cycle.
Hence the projection map $j:r\rightarrow R$ induces a map, $j_*$, of
absolute bulk homology classes to
relative homology classes:
$$
j_*:\qquad H_{m-p-1}(\M;\Z)\longrightarrow H_{m-p-1}(\M,\B;\Z).
\eqnu{homj}
$$\label\homj{\num}\par\noindent
This is actually a homomorphism. Its kernel  consists of bulk cycles,
$r$ for which $R$ is relatively trivial,
so $r=(\partial C+\beta)+\alpha=\partial C+\gamma$,
where $\gamma\in\B$. Since $r$ is closed, so is $\gamma$.
Thus $\gamma$ is a cycle in the boundary and the kernel can be denoted
$$
K_{m-p-1}(\M,\Z)=\{\hbox{classes of bulk cycle homologous to cycles in
}\B\}.
\eqnu{km}
$$\label\km{\num}\par\noindent
On the other hand the image of $j_*$ consists of classes of relative cycle
whose boundary is the boundary of a chain within $\B$.
This coincides with the subgroup $K_{m-p-1}(\M,\B;\Z)$
already defined as being the kernel of $\partial_*$
in the previous section, \kb.

Putting together $j_*$ and $\partial_*$ as two successive homomorphisms:
$$
H_{m-p-1}(\M;\Z)\quad{\buildrel j_*\over\longrightarrow}\quad
H_{m-p-1}(\M,\B;\Z)\quad{\buildrel\partial_*\over
\longrightarrow}\quad H_{m-p-2}(\B;\Z),
$$
we see that this sequence is
exact at $H_{m-p-1}(\M,\B;\Z)$ as $K_{m-p-1}(\M,\B;\Z)$ is both the the
image of $j_*$ and the kernel of $\partial_*$.
This is a short segment of the exact sequence of relative homology
alluded to in the introduction and more segments will be seen when
magnetic fluxes are considered next. The complete exact sequence will
be presented in later sections. A textbook presentation can be found
in [Massey 1991].

Of course, as we saw at the start, all electric charges vanish that are
integrals over cycles in $H_{m-p-1}(\M;\Z)$. This agrees with the fact
already found above
that they also vanish on $K_{m-p-1}(\M,\B;\Z)$, which is the
image of the former group under the action of $j_*$.

\section{Action principle for p-branes and magnetic flux quantisation}

The standard (naive) expression for the term in the action describing the
interaction of the field strength, $F$ with the current, $j$,
that is its source, according to \maxa\ and \maxb, is
$$
\int_{\M}A\wedge *j.\eqnu{intera}$$\label\intera{\num}\par\noindent
Naively, this term is gauge invariant on its own with respect
to the transformation \gauge\ given that the electric current, $j$,
is conserved, \cc, and localised, \twob.
Ideally the current should be expressible in terms of
quantum mechanical wave functions for the matter but it is only really
understood how to do this when $p=0$ so that the branes are point
particles.

By default, the only accepted way to proceed is to adopt
the classical geometric picture described in the introduction.
The evolution of the $p$-brane in space-time is specified by its
world-volume, $w$, an absolute bulk $(p+1)$-cycle on $\M$.
Then the action term \intera\ takes the form \inter\ mentioned at the
start.

Because we already know the classical equations of motion in the Maxwell
form
\maxa\ and \maxb, the detailed form of the action is only really relevant
in the quantum theory. In that context, the expressions \inter\ and
\intera\ are equally problematical (which explains the use of the words
\lq\lq schematic or naive'') as they
involve the gauge potential, $A$, which is only defined locally,
whilst the integration extends globally over all of space-time, $\M$.
Consequently, in a topologically
complicated space-time such as the one being imagined, there are
problems in patching together this expression in overlapping
neighbourhoods of
space-time. Fortunately it is the exponentiated action
$\e^{\im q\int_{w}A/\hbar}$ that enters the Feynman action principle
and this is more amenable. One needs to know how this
phase alters when $w$  is altered by a boundary. That is tantamount to
requiring that the phase  has a meaning when $w$ is a
boundary of a bulk chain. This can be done provided
the background field strength $F$
satisfies the Dirac quantisation conditions for all
magnetic fluxes through bulk $(p+2)$-cycles:
$$
\Phi_M(v)=\int_{v}F\in {2\pi\hbar\over q}\,\Z,\qquad \partial
v=0.\eqnu{mflux}
$$\label\mflux{\num}\par\noindent
As $dF=0$ these fluxes are defined on the classes of the absolute
homology of space-time $\M$, forming the group $H_{p+2}(\M,\Z)$, or
more precisely the free part of this, $F_{p+2}(\M;\Z)$, a lattice of
dimension $b_{p+2}(\M)$.

A parenthetic remark concerning this quantisation condition \mflux\ is
that it
is known not really to be correct when wave functions are considered,
as is, so far, only possible when $p$ vanishes
and the brane is therefore a point
particle. Then there is a possibility of fractional quantisation
conditions when the wave function is of a spinor nature (involving
half-integers instead of integers). The precise
rule is easy to state when $m=4$ [Alvarez and Olive 2000].

By this stage of the argument it has become established that, as claimed
in the introduction, there is a connection between the physical notions of
electric charge, electric flux and magnetic flux of a $p$-brane and
mathematical notions of relative homology, absolute boundary homology
and absolute bulk homology
and more precisely, with the free parts of the abelian homology groups,
$H_{m-p-1}(\M,\B;\Z)$, $H_{m-p-2}(\B;\Z)$ and $H_{p+2}(\M;\Z)$,
respectively.
But there is a more detailed structure connected to the subgroups
$K_*$ of $H_*$, for short, that plays a role in the exact sequence
of relative homology and moreover possesses a physical relevance.
Let us illustrate this last point by investigating magnetic fluxes
through $\B$ cycles with a view to comparing electric and magnetic
fluxes. Later on we shall see how this comparison will indicate
a generalised dyonic phenomenon that is possibly related to the
Zwanziger-Schwinger quantisation condition [Zwanziger 1968, Schwinger
1969].

Magnetic fluxes can already be defined for cycles in the boundary,
$\B$, rather than in the bulk, $\M$, but nothing appears to be gained
by this as cycles
in $\B$ are automatically cycles in $\M$ but may become boundaries
of bulk chains when regarded as $\M$-cycles and hence homologically
trivial in the bulk. Associated with this idea is the
inclusion map, $i$, which induces the homomorphism:
$$
i_*:\qquad H_{p+2}(\B;\Z)\longrightarrow H_{p+2}(\M;\Z),
\eqnu{homi}
$$\label\homi{\num}\par\noindent
with kernel consisting of the classes of cycle just mentioned
that become boundaries. This is precisely the subgroup $K_{p+2}(\B;\Z)$
of the type met before, \kb, (with $p+2$ replaced by $m-p-2$),
as the image of the homomorphism, $\partial_*$, \homd, induced by the
boundary
operator and met before in the comparison of electric charges and fluxes.
The image of this homomorphism is clearly given by classes of bulk cycle
homologous to a cycle in the boundary and these precisely form
the subgroup $K_{p+2}(\M;\Z)$, \km, already met as the kernel of
the homomorphism $j_*$, \homj, (again with $p+2$ replaced by $m-p-2$).

All magnetic fluxes on cycles of $K_{p+2}(\B;\Z)$ vanish, as
$$
\int_{\phi}F=\int_{\partial S}F=\int_S dF=0,
$$
by \kb, Stokes' theorem and \maxa, corresponding to the fact that these
cycles are trivial as bulk cycles. So the only non-trivial magnetic fluxes
through boundary cycles correspond to the $b_{p+2}(\B)-s_{p+2}(\B)$ cosets
of $K_{p+2}(\B;\Z)$ in $H_{p+2}(\B;\Z)$. These observations will become
more
interesting when we are able to compare them with the corresponding
properties of electric fluxes through boundary cycles later on.

Thus we have two more examples of a coincidence between images and kernels
of different homomorphisms. This phenomenon is part of the exact
sequence of relative homology mentioned in the introduction,
an important pattern that has been emerging gradually
and will be elaborated now.

\section{The exact sequence of relative homology of space-time}

Our study within a general setting of the physical concepts of
electric charge, electric flux and magnetic flux has revealed
how these are described as integrals over cycles in space-time that
are respectively relative, boundary and bulk type and unchanged by
the appropriate homologies. So they are certainly classified
by the corresponding homology groups $H_{m-p-1}(\M,\B;\Z)$,
$H_{m-p-2}(\B;\Z)$
and $H_{p+2}(\M,\Z)$, when $p$-branes are considered.

We  have also met three different types of homomorphism between
the three types of homology group, denoted $i_*$, $j_*$ and
$\partial_*$, and illustrated by \homi, \homj\ and \homd.
Associated with all of these is an image and kernel which
is always a very specific subgroup of the relevant
homology group, illustrated by \kmb, \kb\ and \km.

If $p$ is allowed to run over all the values compatible with possible
$p$-branes in the given background space-time $\M$, the set of all
possible homology groups can be arranged as an ordered sequence
with homomorphisms of one
or other of the above three types relating each successive pair:-

$$
...\buildrel\partial_*\over\longrightarrow H_{m-p-1}(\B)
\buildrel i_*\over\longrightarrow H_{m-p-1}(\M) \buildrel
j_*\over\longrightarrow H_{m-p-1}(\M,\B)
\buildrel\partial_*\over\longrightarrow
H_{m-p-2}(\B) \buildrel i_*\over\longrightarrow  ...
\eqnu{homexact}
$$\label\homexact{\num}\par

This is the exact sequence of relative homology  well known to pure
mathematicians in the context of algebraic topology,
and more careful and detailed treatments can be found in various textbooks.
The notation has been compressed by omitting reference to the integers
$\Z$.

Assuming space-time, $\M$, is connected, this exact sequence of abelian
groups
starts and finishes with the trivial group, written as
$1$ in multiplicative notation:

$$
1\rightarrow H_m(\M,\B)\rightarrow H_{m-1}(\B) \rightarrow H_{m-1}(\M)
\rightarrow H_{m-1}(\M,\B)\rightarrow\dots
$$
and
$$
\dots H_1(\B)\rightarrow H_1(\M)\rightarrow H_1(\M,\B)\rightarrow
H_0(\B)\rightarrow H_0(\M)\rightarrow 1.
$$

Thus, besides the two trivial terms terminating the exact sequence, there
are $3m$ terms. From the sequence it is now possible to evaluate in terms
of the Betti numbers the numbers $s_q(\B)$, $s_q(\M)$ and $s_q(\M.\B)$ that
are the dimensions of the free parts of the kernels \kb, \km\ and \kmb\ and
entered the counts of the various  charges and fluxes.

The exact sequence \homexact\ implies a similar but simpler exact
sequence for the free parts of the homology groups
(obtained by dividing out the torsion subgroup). Working over
real coefficients rather than integers yields an exact sequence
of vector spaces
with dimensions given by the Betti numbers and linked by linear maps
replacing the group homomorphisms. To understand what happens consider
such a sequence in simplified notation:
$$
1\rightarrow V_0 \rightarrow V_1\rightarrow V_2 \rightarrow V_3
\rightarrow\dots V_N\rightarrow 1
\eqnu{vseq}
$$\label\vseq{\num}\par\noindent

If $K_n\subset V_n$ is the kernel/image, then,
by exactness $K_n\equiv V_{n-1}/K_{n-1}$ (retaining multiplicative
notation).
So repeating
$$
K_n=V_{n-1}/V_{n-2}/V_{n-3}/\dots /V_1/V_0/1,
\eqnu{ker}
$$\label\ker{\num}\par\noindent
and taking dimensions,
$$
s_n=\hbox{dim}K_n=b_{n-1}-b_{n-2}\dots (-1)^{n+1}b_0=b_n-b_{n+1}\dots
(-1)^{N-n}b_N,
\eqnu{dimk}
$$\label\dimk{\num}\par\noindent
using the fact that $s_{N+1}$, which equals the alternating sum of
all the Betti numbers, vanishes.

Applying these formulae to the exact sequence of relative homology,
\homexact,
yields
$$s_n(\M)=b_n(\M)-b_{n}(\M,\B)+b_{n-1}(\B)-b_{n-1}(\M)\dots,$$
$$s_n(\M,\B)=b_n(\M,\B)-b_{n-1}(\B)+b_{n-1}(\M)-b_{n-1}(\M,\B)\dots,$$
$$s_n(\B)=b_n(\B)-b_n(\M)+b_n(\M,\B)-b_{n-1}(\B)\dots,$$
showing how the count of electric charges, \chcount, depends on
the topology of space-time, $\M$.

It is familiar that in the the understanding of electro-magnetic
duality on closed
space-time manifolds, $\M$,  a property known as Poincar\'e duality
is important.
There is an analogous property for manifolds with boundary that will
play an important role in the present context. This is known
as Poincar\'e-Lefschetz duality and a short explanation follows.

Corresponding to the integer homology groups already defined it is possible
to define integer cohomology groups denoted $H^q(\M;\Z)$ and so on.
There is also an exact sequence
of homomorphisms linking these in the sense of ascending superscript:
$$
\dots\rightarrow H^{p}(\B)\rightarrow H^{p+1}(\M,\B)\rightarrow
H^{p+1}(\M)\rightarrow
H^{p+1}(\B)\rightarrow\dots
\eqnu{cohomex}
$$\label\cohomex{\num}\par\noindent

The statement of Poincar\'e-Lefschetz duality is that the corresponding
terms
in the two exact sequences \homexact\ and \cohomex\ are isomorphic as
groups.
So
$$
H^{p+1}(\M,\B;\Z)\equiv H_{m-p-1}(\M;\Z), \qquad H^{p+1}(\M;\Z)\equiv
H_{m-p-1}(\M,\B;\Z),
\eqnu{pla}
$$\label\pla{\num}\par\noindent
and
$$
H^p(\B;\Z)\equiv H_{m-p-1}(\B;\Z).
\eqnu{plb}
$$\label\plb{\num}\par\noindent
The last isomorphism is simply Poincar\'e duality for the boundary, $\B$,
which is automatically a closed manifold of dimension $m-1$. Notice
how the superscripts and subscripts in an isomorphism are always
complementary in the sense of summing to the dimension
of the relevant manifold and how \pla\ relates relative topology to
absolute
topology in the bulk.

There is yet another relation between homology and cohomology that results
from the universal coefficient theorem by considering the coefficients
to be real numbers rather than integers. The resultant groups are 
simply the vector spaces, with dimension equal to the Betti number, spanned
by the lattices given by the free parts
of the integer groups as previously mentioned. Then a homology group
of given suffix
and type is the dual of the cohomology group of
corresponding superscript and type:
$$
H^q(\M;\R)=H_q(\M;\R)^*,\quad H^q(\M,\B;\R)=H_q(\M,\B;\R)^*,\quad
H^q(\B;\R)=H_q(\B;\R)^*.
\eqnu{dual}
$$\label\dual{\num}\par\noindent
By means of these and the Poincar\'e-Lefschetz duality
relations \pla\ and \plb, the cohomology groups can be eliminated to yield
the following relations between homology groups:
$$
H_q(\M;\R)=H_{m-q}(\M,\B;\R)^*\qquad\hbox{and}\qquad
H_q(\B;\R)=H_{m-q-1}(\B;\R)^*.
\eqnu{pliso}
$$\label\pliso{\num}\par\noindent
Because the Betti numbers are  the dimensions of these real vector spaces,
particular consequences are the following equalities:
$$
b_q(\M)=b_{m-q}(\M,\B)\qquad\hbox{ and }\qquad b_q(\B)=b_{m-q-1}(\B).
\eqnu{plbet}
$$\label\plbet{\num}\par\noindent
The corresponding duality relations for the dimensions on the image/kernels
of the exact sequence, the numbers $s_q(\M)$, $s_q(\B)$ and $s_q(\M,\B)$,
will be important and are easily obtained  by recognising that in the
simplified notation for the exact sequence, \vseq, $V_n=V_{N-n}^*$.
So $b_n=b_{N-n}$ and hence by \dimk,
$$
s_n=s_{N+1-n}.
$$
As a consequence
$$
\hbox{dim}\,V_n=b_n=s_n+s_{n+1}=\hbox{dim}K_n+\hbox{dim}(V/K)_n
$$
and
$$
b_{N-n}=
\hbox{dim}\,V_{N-n}=s_{N+1-n}+s_{N-n}=\hbox{dim}(V/K)_{N-n}
+\hbox{dim}\,K_{N-n}.
$$
This means the dimensions of the two complementary subspaces of $V$,
namely $K$ and $V/K$ interchange under duality, $N\leftrightarrow N-n$.

In particular
$$
s_{m-p-1}(\M,\B)=b_{p+1}(\M)-s_{p+1}(\M) \hbox{ and }
s_{p+1}(\M)=b_{m-p-1}(\M,\B)-s_{m-p-1}(\M,\B).
\eqnu{kdu}
$$\label\kdu{\num}\par\noindent
In fact, by \plbet\ these two equations are the same as each other.

\section{Electric charges as intersection numbers}

With this information we are now well
prepared to consider the physical question as to the possible
numerical values of the generalised electric charges \charge. Given a
suitable
expression for the conserved,
localised electric current, $j$, the charges are evidently determined
without
recourse to  Maxwell's equations \maxa\ and \maxb.
Hence in this calculation these equations can be temporarily
renounced, provided it is remembered that their reinstatement will
reduce the number of independent electric charges, as explained in
section 3.
We shall defer this reinstatement and the detailed
understanding of the issues it raises until the following section.

Just as in the discussion of magnetic fluxes and their quantisation
in section 6, we shall have to resort to the
geometrical picture of a brane world-volume, as this will give us tractable
form for the current. This is found by equating \inter\ and \intera, the
two
versions of the term in the action responsible for the brane coupling to
the
gauge potential:
$$
q\int_w A=\int_{\M}A\wedge*j.
\eqnu{interc}
$$\label\interc{\num}\par\noindent
Now $A$ is taken to be an arbitrary $(p+1)$-form on $\M$, so it follows
that
$$
*j=q\mu(w),
\eqnu{derham}
$$\label\derham{\num}\par\noindent
where $\mu(w)$ is a singular $(m-p-1)$-form involving a product
of the same number of Dirac $\delta$-functions  with
support on the absolute $(p+1)$-cycle
$w$ and differentials in the variables transverse to it. It follows that
its restriction to $\B$ vanishes,
as it should \twob. In the Appendix it will be shown  to be closed as well.

Inserting the $p$-brane current \derham\ into the electric charge \charge\
yields
$$
Q(w;S)=q\int_{S}\mu(w).
$$
This is invariant under relative homologies of $S$ according
to the work of section 3. Now consider a bulk homology of the world volume,
$w$, $w\rightarrow w'=w+\partial a$. By linearity $\mu(w')=
\mu(w)+\mu(\partial a)$. As discussed in the Appendix,  $\mu(a)$ exists
for a bulk chain, $a$
(and now involves step functions as well as Dirac-delta functions)
and, moreover, obeys $d\mu(a)=\mu(\partial a)$, up to a sign.
Hence the change in the electric charge, $Q(w;S)$, due to this homology is
$$
Q(w'-w;S)=q\int_S\mu(\partial a)=q\int_Sd\mu(a)=q\int_{\partial
S}\mu(a)=\int_{\partial S}\mu(a)\atbound=0
$$
So $Q(w;S)$ is defined on the homology classes
$H_{m-p-1}(\M,\B;\Z)\times H_{p+1}(\M;\Z)$, or rather on the
corresponding product of free parts. So it
can be assumed that
the relative cycle $S$ intersects the absolute bulk cycle of complementary
dimension, $w$, at discrete points. Then the integral for the electric
charge is recognised as [Henneaux and Teitelboim 86]
$$
Q(w;S)=qI(w,S)
$$
where $I(w,S)$ denotes the intersection number of  the absolute bulk
cycle $w$ with  the relative cycle $S$, being the algebraic sum of the
number of these points, taking into account signs due to relative
orientation.
This intersection number possesses a number of mathematical properties that
are important for the physical interpretation of this result that we shall
now describe.

Choose bases ${S_j}$ and ${w_i}$ in the lattices $F_{m-p-1}(\M,\B;\Z)$ and
$F_{p+1}(\M;\Z)$ that are the free parts of the two relevant homology
groups.
Then all intersection numbers are specified by knowledge of the matrix
$$
I(w_i,S_j)=I_{ij}\qquad \in\Z
\eqnu{intersection}
$$\label\intersection{\num}\par\noindent
This intersection matrix, $I$, has  $b_{p+1}(\M)$  rows and
$b_{m-p-1}(\M,\B)$  columns and hence is square, by \plbet. Yet another
consequence of Poincar\'e-Lefschetz duality is that this matrix $I$
is unimodular:
$$
\hbox{det }I=\pm 1.
\eqnu{unim}
$$\label\unim{\num}\par\noindent
Putting these results together it follows that all electric charges
are quantised:
$$
Q(S)\in q\Z
\eqnu{chqu}
$$\label\chqu{\num}\par\noindent
as integral multiples of the coupling constant, $q$, that enters the
action.
Thus any electric charge paired with any magnetic flux satisfies the Dirac
quantisation condition:
$$
Q(S)\Phi_M(v)\in 2\pi\hbar\Z.
\eqnu{dirac}
$$\label\dirac{\num}\par\noindent
If Maxwell's equations are permitted, then it follows that certain
of the electric fluxes are indeed quantised, namely those obtained by
integrating over boundary cycles within $K_{m-p-2}(\B;\Z)$, as these fluxes
are equal to electric charges.

On the other hand, there is no reason to believe that the remaining
electric fluxes are quantised, and we shall return to some comments on this
later. All that can be said as a result of \chqu\ is that, for the
latter fluxes, the quantities $\hbox{exp}({{2\pi i \Phi_E(v)\over
q}})$ are well defined on the cosets $(H/K)_{m-p-2}(\B;\Z)$.

The physical consequence of $I$ being unimodular is that a configuration of
the brane-world-volume can be found that realises any assignment of charges
satisfying \chqu.

Finally let us comment on the connection between the physical arguments
of this section and the mathematical arguments of the preceding one,
outlining Poincar\'e-Lefschetz duality. By current conservation, \cc,
the dual current, $*j$,  is a closed $(m-p-1)$-form, that, in addition,
has vanishing restriction on the boundary of space-time, \twob. This means
that it is a relative $(m-p-1)$-cocycle in the sense of de Rham cohomology,
and so defines
a class of $H_{\rm{de~Rham}}^{m-p-1}(\M,\B;\R)$. The same is therefore
true of $\mu(w)$ by \derham, which therefore provides a map from the
homology
class of the world-volume, $w$, $H_{p+1}(\M;\Z)$ to
$H_{\rm{de~Rham}}^{m-p-1}(\M,\B;\R)$.
This is part of the Poincar\'e-Lefschetz isomorphism, \pla.
It is possible to develop this line of thought and this is done in the
Appendix.

\section{Maxwell's equations and the intersection matrix}
Two (correct) arguments have been developed in this paper that apparently
lead to a contradiction. We shall now explain what this is, and how it
is resolved by finding that the intersection matrix, $I$, \intersection,
has further detailed properties, hitherto unexpected.

In section 4 it was shown that the effect of Maxwell's equations is
to force all electric charges, $Q(S)$, to vanish when the relative cycle
over which they are integrated, $S$, belongs to $K_{m-p-1}(\M,\B;\Z)$. Yet
according to the preceding section, irrespective of Maxwell's equations,
the electric charge due to a brane configuration with world-volume, $w$ was
seen to be
proportional to the intersection number of $w$ with $S$. The apparent
contradiction arises from the fact that the intersection matrix is
non-singular, as a consequence of its being unimodular \unim.

To make this clearer, it is natural to partition the intersection matrix
in a way that distinguishes each  kernel $K$ within each $H$ from
the cosets $H/K$. This is done by choosing the basis $\{S_j\}$ so that
the first $s_{m-p-1}(\M,\B)$ elements form a basis of $K_{m-p-1}(\M,\B)$
while
the remainder refer to the cosets $(H/K)_{m-p-1}(\M,\B)$. The basis
$\{w_i\}$ is chosen so that the last $s_{p+1}(\M)$ elements form a basis
for
$K_{p+1}(\M)$ while the remainder refer to the cosets. Corresponding
to this, the intersection matrix, \intersection, is written in the block
form
$$
I(w,S)=\bordermatrix{&{\scriptscriptstyle K(\M,\B)}& &
{\scriptscriptstyle (H/K)(\M,\B )}\cr
{\scriptscriptstyle (H/K)(\M)}&A& &Y\cr &&&\cr&&&\cr{\scriptscriptstyle
K(\M )}&X& &B\cr}.
\eqnu{intermat}
$$\label\intermat{\num}\par\noindent
That electric charges associated with $K_{m-p-1}(\M,\B)$ all vanish
seems to imply that the submatrices $A$ and $X$ vanish,
apparently contradicting the fact that
the overall matrix has determinant equal to $\pm1$.

But this is not a correct interpretation of what has been shown.
The correct interpretation is that  the absolute bulk homology classes
of the brane world-volume, $w$, that yield non-zero charges associated
with relative cycles of  homology belonging
to the subgroup $K_{m-p-1}(\M,\B;\Z)$
are all forbidden because Maxwell's equations cannot
then be integrated to yield
field strengths, given the corresponding currents \derham. Thus the only
permitted homology classes of world-volume are those whose intersection
number with all elements of $K_{m-p-1}(\M,\B;\Z)$ vanish. These classes
should form a subgroup and it natural to anticipate that this be provided
by the kernel $K_{p+1}(\M;\Z)$. The condition for this is that
the submatrix $X$ in \intermat\ vanish. This is perfectly consistent
with the unimodularity of I, \unim, since, by \kdu, the consequences
of Poincar\'e-Lefschetz duality for the kernels, the block
diagonal  submatrices $A$ and $B$ are both square. Consequently
$$
\pm 1=\hbox{det }I=\hbox{det }A\,\,\hbox{det }B,
$$
implying that the block diagonal submatrices $A$ and $B$, possessing
integer
entries, are both unimodular too.

Thus it is the submatrix $B$ that gives the physical charges, $Q(S)$, for
$S$ in the coset $(H/K)_{m-p-1}(\M,\B;\Z)$
as it determines the intersection numbers between these relative classes
and the permitted homology classes of brane world-volume.
According to \km, (with $m-p-1$
replaced by $p+1$) these permitted world-volumes are those homologous to
cycles in the boundary of space-time, $\B$. It is remarkable that such a
selection rule on brane configurations can be derived without recourse to
any equations of motion for the brane degrees of freedom.

Since the submatrix, $B$, that determines the physical charges, is
unimodular,
the previous conclusion there exist brane configurations realising
any assignment of quantised charges, \chqu, holds good even when the
selection rule is taken into account. What remains is
to provide an independent check that the
block submatrix $X$ in \intermat\ vanishes.

This is a geometrical condition that should hold for any
background space-time, $\M$,  with boundary $\B$, and it can be
rewritten as:
$$
I(K_{p+1}(\M),K_{m-p-1}(\M,\B))=0.
\eqnu{izero}
$$\label\izero{\num}\par\noindent
This vanishing theorem will be demonstrated in the next section
using some results developed in the Appendix.

\section{De Rham cohomology, field strengths and currents}
The argument will be interesting as it brings into play further parallels
between physical and mathematical concepts and
sheds light on the more
abstract  ideas involving  the two related exact sequences mentioned
previously
and to be elaborated below.

We can no longer avoid describing de Rham cohomology which deals with the
exterior derivative, $d$, of differential forms
(such as the field strengths and currents we have been talking about).
We  have to explain the three types of cohomology group that arise,
given a manifold with boundary; how they can arranged in an exact sequence,
and how that exact sequence is related to the one for homology groups
already explained.

A real $q$-form, $\omega$, on $\M$ is an absolute bulk cocycle if
it is coclosed, $d\omega=0$. Two such cocycles are absolutely
cohomologous in the bulk if
$$
H^q(\M;\R):\qquad \qquad\omega'\sim \omega\quad
\Longleftrightarrow\quad \omega'=\omega+d\alpha.
\eqnu{mcoh}
$$\label\mcoh{\num}\par\noindent
As indicated, these form equivalences classes which constitute elements
of the absolute bulk homology group $H^q(M)$ (in the sense of de
Rham). The groups are abelian since composition is by
addition. Actually these groups are real vector spaces since they are
closed under multiplication by real numbers.

Essentially the same concepts can be applied to $q$-forms, $\phi$, on the
boundary, $\B$. $\phi$ is a boundary cocycle if it exists on $\B$ and is
coclosed there. Two such cocycles are absolutely cohomologous
on the boundary if
$$
H^q(\B;\R):\qquad\qquad \phi'\sim\phi\quad \Longleftrightarrow \quad\phi'=
\phi+d\beta.
\eqnu{bcoh}
$$\label\bcoh{\num}\par\noindent
Again these are equivalence relations whose classes form the group
indicated.

The third and last concept is that of a relative cocycle, $\eta$, which
is defined in the bulk, on $\M$, is coclosed there, $d\eta=0$ and
has vanishing restriction to the boundary, $\eta\atbound=0$. Two
such cocycles are relatively cohomologous if they differ by a coexact
form $d\alpha$ with the property that the restriction of $\alpha$ to
the boundary is coexact there.
$$
H^q(\M,\B;\R):\qquad\qquad \eta'\sim\eta\quad \Longleftrightarrow\quad
\eta'=\eta+d\alpha, \quad \alpha\atbound=d\beta.
\eqnu{rcoh}
$$\label\rcoh{\num}\par\noindent
Again these are equivalence relations whose classes form the group
indicated.
In each case the cohomology relation preserves the appropriate
coclosure property. Physical examples are provided by the field
strength, $F$,
which is an absolute bulk
$(p+2)$-cocycle, $*F\atbound$, which is an absolute boundary
$(m-p-2)$-cocycle
and the dual current, $*j$, which is a relative $(m-p-1)$-cocycle.

Thus there are three types of cohomology matching the three types of
homology
already explained. Furthermore they both  exist for a range of values of
the integer $q$ specifying the dimension of the cycle or the degree of
the form as appropriate. When taken over real numbers, homology
and cohomology groups of matching
type and integer $q$ are related in a nice way, as dual vector spaces,
see \dual. To understand
the first example of these relations, let $\omega$ be a $q$-cocycle and
$v$ a
$q$-cycle, both in the absolute bulk sense, and consider
$$
\int_v\,\omega\in \R.
$$
This integral enjoys a number of properties:

\noindent 1) It is invariant under the appropriate homologies
of $v$, $v\rightarrow v'=v+\partial a$,
and cohomologies of $\omega$, \mcoh .

\noindent 2) It is linear in $v$ and $\omega$ separately and hence provides
a real bilinear form.

\noindent 3) It is nonsingular; that is there is no nontrivial class
of either type such
that the integral vanishes for all   classes of the other type.

The first two properties are easy to check but the third, nonsingularity,
is quoted as a known theorem (of de Rham).
Of course the magnetic flux \mflux\ already defined is an example
of such an integral.
Precisely analogous constructions work for the other two types
of homology/cohomology and yield the remaining duality
relations \dual. Physical examples of these
integrals are provided by electric charge, \charge, and electric flux,
\eflux,
involving relative and boundary homology/cohomology respectively.

Space-time, $\M$, is itself a relative $m$-cycle and hence it is
appropriate
to integrate relative $m$-cocycles over it. The wedge product
$\eta\wedge\omega$ is such a cycle
if $\eta$ and $\omega$ are respectively relative and   absolute bulk
cocycles
of complementary degree (summing to $m$). So it is natural to consider
$$
\int_{\M}\eta\wedge\omega\in \R.
$$
This integral is

\noindent (1) invariant under the appropriate cohomologies of $\omega$
and $\eta$, \mcoh\ and \rcoh, 

\noindent (2) bilinear in $\omega$ and $\eta$

\noindent (3) nonsingular.

As a consequence  there results the duality relation
$$
H^q(\M;\R)=H^{m-q}(\M,\B;\R)^*
$$
which, when combined with the previous duality relations \dual, implies
$$
H^q(\M;\R)=H_{m-q}(\M,\B;\R)\quad\hbox{and}\quad H^q(\M,\B;\R)=
H_{m-q}(\M;\R),
$$
a weak version of Poincar\'e-Lefschetz duality, \pla\ (weak because
it is over the reals rather than the integers). A weak version (over the
reals)
of the  similar relation for the boundary, \plb, can likewise be checked.

These results are sufficient to show that there exists an exact sequence of
de Rham cohomology groups but it is worth demonstrating this explicitly
in order to find precise definitions of the common kernel/image
subgroups of
these groups.

Relative cocycles are automatically absolute cocycles in the bulk too and
this leads to the homomorphism
$$
j^*:\qquad H^q(\M,\B)\rightarrow H^q(\M).
\eqnu{jhom}
$$\label\jhom{\num}\par\noindent
The kernel of $j^*$ is made up of the elements that are trivial in
$H^q(\M)$:
$$
K^q(\M,\B)=\{\hbox{classes of }H^q(\M,\B)\hbox{ satisfying
}\eta=d\alpha,\quad
d\alpha\atbound=0\},
\eqnu{mbk}
$$\label\mbk{\num}\par\noindent
while the image appears to consist of elements of $H^q(\M)$ with
$\omega\atbound$ vanishing.

Absolute cocycles in the bulk automatically yield cocycles in
the boundary when restricted to it. So $\omega\rightarrow\omega\atbound$
yields the homomorphism
$$
i^*:\qquad H^q(\M)\rightarrow H^q(\B)
\eqnu{ihom}
$$\label\ihom{\num}\par\noindent
with kernel
$$
K^q(\M)=\{\hbox{classes of } H^q(\M) \hbox{ with }\omega\atbound
\hbox{ coexact}\}
\eqnu{mk}
$$\label\mk{\num}\par\noindent
This obviously includes the image of $j^*$ and tallies
after applying bulk cohomologies \mcoh.
The image of $i^*$ will be specified below.

Given a coclosed form, $\beta_0$, on the boundary, $d\beta_0=0$ on $\B$,
there
is a way to find a closed form $\eta_\beta$, of one degree
higher on the bulk whose restriction
to the boundary automatically vanishes so that it is relatively coclosed.
Although the procedure is not unique, the degree of ambiguity
lies in a single relative cohomology class and so the procedure leads
to a homomorphism, known as the Bockstein homomomorphism:
$$
d^*:\qquad H^q(\B)\quad
\rightarrow\quad H^{q+1}(\M,\B)
\eqnu{dhom}
$$\label\dhom{\num}\par\noindent
Let $\beta$ denote an extension of $\beta_0$ from the boundary, that is,
a form on $\M$, not necessarily closed, satisfying $\beta\atbound=\beta_0$.
Then, if $\eta_{\beta}=d\beta$, $d\eta_{\beta}=0$ and
$\eta_{\beta}\atbound=d\beta\atbound=d\beta_0=0$ and
so $\eta_{\beta}$ is
a relative cocycle. Consider now $\beta_0$ and $\beta_0'$, forms which
are cohomologous in $H^q(\B)$, \bcoh, so $\beta_0'=\beta_0=d\alpha$,
(on $\B$).
If they have extensions $\beta$ and $\beta'$, respectively to the bulk
$$
\eta_{\beta'}-\eta_{\beta}=d(\beta'-\beta)
\qquad\hbox{and}\qquad (\beta'-\beta)\atbound=d\alpha,
$$
which means $\eta_{\beta'}$ and $\eta_{\beta}$ are
relatively cohomologous, \rcoh, as desired. In particular, 
this applies to the ambiguity arising when $\beta'$ and $\beta$
are different extensions of the same $\beta_0$.

The image of $d^*$ is obviously given by increasing $q$ by unity in \mbk,
originally the kernel of $j^*$ but the kernel of $d^*$ is trickier.
Obviously $\eta_{\beta}$ is
trivial in relative cohomology \rcoh\ whenever $\beta_0$ is coexact but
this
means it is trivial in $H^q(\B)$. But $\eta_{\beta}$ is also trivial
if it vanishes,
that is if $\beta_0$ extends to a form $\beta$ in the bulk which is still
coclosed. Thus
$$
K^q(\B)=\{\hbox{classes of }H^q(\B)
\hbox{ extending to coclosed forms on }\M\}.
\eqnu{bk}
$$\label\bk{\num}\par\noindent
This is also the image of $i^*$. Thus we have a series of identifications
of images and kernels and the results can be all assembled in the
following grand diagram:
$$
\matrix{ ..\buildrel i^*\over\rightarrow &
\!\!H^{p}(\B)\!\!&{\buildrel d^*\over\rightarrow} &\!\!H^{p+1}(\M,\B)\!\!&
\buildrel j^*\over\rightarrow &\!\!H^{p+1}(\M)\!\!&\buildrel i^*\over
\rightarrow &\!\!H^{p+1}(\B)\!\!&\buildrel d^*\over\rightarrow .. \cr
\noalign{\smallskip}
&\big\updownarrow& &\big\updownarrow & &\big\updownarrow&&
\big\updownarrow&\cr..\buildrel \partial_*\over\rightarrow &
\!\!H_{m-p-1}(\B)\!\!&\buildrel i_*\over\rightarrow &\!\!H_{m-p-1}(\M)\!\!&
\buildrel j_*\over\rightarrow &\!\!H_{m-p-1}(\M,\B)\!\!&\buildrel
\partial_*\over\rightarrow&\!\!H_{m-p-2}(\B)\!\!&\buildrel i_*\over
\rightarrow.. \cr}
\eqnu{gseq}
$$\label\gseq{\num}\par\noindent
The upper sequence is composed of the homomorphisms of the de Rham
cohomology groups just described.
It is exact because at each stage the kernels
and images coincide as was just explained. The lower sequence is
the exact sequence of homology \homexact\ explained in previous
sections whilst the vertical arrows indicate the Poincar\'e-Lefschetz
isomorphisms \pla\ and \plb. The most powerful version of this
diagram refers to groups taken over the integers, $\Z$, but for some parts
of the diagram we have
only given arguments establishing a weaker version, over the reals, $\R$.

By \ker, a consequence of exactness, the kernel subgroups
of the pairs of groups related by the Poincar\'e-Lefschetz isomorphism
are themselves isomorphic. This suggests that
the result we wish to prove, \izero, is equivalent to its
cohomological counterpart:
$$
\int_{\M}\eta\wedge\omega =0 \qquad\qquad\hbox{if}\quad \quad\eta\in
K^{m-q}(\M,\B;\R)\quad\hbox{ and } \omega\in
K^q(\M;\R).
\eqnu{van}
$$\label\van{\num}\par\noindent
But this is quite easy to prove using the results above, as we now see.
By \mbk\ the integral equals $\int_{\M}d\alpha\wedge \omega=
\int_{\M}d(\alpha\wedge\omega)$ as $d\omega$ vanishes by \mk.
By Stokes' theorem on $\M$ the integral equals
$\int_{\B}\alpha\wedge\omega=
\int_{\B}\alpha\atbound\wedge\omega\atbound=
\int_{\B}\alpha\wedge d\gamma$ as $\omega\atbound$ is coexact by \mk. But,
by \mbk, $\alpha$ is coclosed so
the integral equals $\int_{\B}d(\alpha\wedge\gamma)=
\int_{\partial\B} \alpha\wedge\gamma=0$, by Stokes' for the
boundary, $\B$,
and the fact that the latter is automatically closed.

Vanishing theorems  analogous to \van\ also apply to the integrals
like $\int_w\omega$
coupling a pair of like homology and cohomology groups. For example,
the electric charge $Q(S)=\int_S*j$ couples the relative homology of
the integration domain, $S$,
$H_{m-p-1}(\M,\B)$ to the relative de Rham cohomology of the dual current,
$*j$, $H^{m-p-1}(\M,\B)$ and vanishes when $S\in K_{m-p-1}(\M,\B)$, \kmb\
and $*j\in K^{m-p-1}(\M,\B)$, \mbk. The latter condition
certainly hold when Maxwell's equation, \maxb, for the field strength,
$F$, holds. This vanishing theorem is then precisely what was proven
in our earlier discussion of electric charges, and that is now seen
to be part of a more general pattern.

The last step is the derivation of the vanishing
theorem \izero\ for the intersection
matrix  from the vanishing theorem for cohomology, \van, proven above,
using the upward arrow in the
Poincar\'e-Lefschetz isomorphism, \gseq. A convenient concrete version
of this
map is provided by the quantity $\mu(w)$
that enters the expression \derham\ for the dual current $*j$
due to a brane whose world-volume is the absolute cycle $w$,
and generalisations of this to be explained in the Appendix.
These maps will provide homomorphisms between the groups
indicated in \gseq\  mapping the appropriate kernel subgroups into each
other.
The desired result  follows by combining these results with the fact that
the intersection number can be written
$$
I(w,S)=\int_{\M}\mu(w)\wedge\mu(S)
$$

\section{Discussion}
Motivated by the physical questions of elucidating and counting
the types of conservation law occurring in the sorts of
generalised Maxwell theories that arise naturally
in string/superstring theories formulated on a fixed background space-time
of possibly complicated topology,
we have been led to a well established
area of pure mathematics. This is the theory of relative
homology/cohomology
associated with the space-time, assumed to have a boundary, and it seems
not
to be  so familiar to physicists despite its evident physical relevance.
Accordingly we have tried to build it up systematically, as guided by
physics,
and in particular, the generalised Maxwell's equations,
and included reasonably self-contained proofs.

Given an understanding of the overall grand mathematical structure,
comprising the two exact sequences
of homology and cohomology and the Poincar\'e-Lefschetz isomorphism
relating them, as depicted by \gseq, and the duality relations, \dual,
indicating a horizontal reflection symmetry of the exact sequences,
it is relatively
easy to explain the relevance to physics. This is what we now do
because of the value of the  new perspectives afforded.

The first step is the recognition that there are precisely three
types of conserved quantity, electric charge, \charge, electric flux,
\eflux\
and magnetic flux, \mflux, and that these are associated with the three
possible types of homology/cohomology,
namely relative, boundary and absolute bulk,
respectively. In fact these conserved quantities are invariant
under the appropriate homologies/cohomologies and, indeed,
constitute non-singular bilinear forms on the free parts of these groups,
thereby being responsible for the duality relations,
\dual, of the exact sequences \gseq.

However this argument makes only partial use of the
generalised Maxwell's equations, \maxa\ and \maxb, and the associated
boundary conditions
\twob\ and \hb. What is used for each conserved quantity in turn is:

\noindent Electric charge, \charge: $d*j=0$ and $*j\atbound=0$

\noindent Electric flux, \eflux: $d\{(*F)\atbound\}=0$

\noindent Magnetic Flux, \mflux: $dF=0$.

With this limited information these three conserved quantities appear
unrelated
to each other and counted by the relevant Betti numbers,
$b_{m-p-1}(\M,\B)$,
$b_{m-p-2}(\B)$ and $b_{p+2}(\M)$ as explained above. The content in
Maxwell's
equations that has not so far been exploited is the
inhomogeneous Maxwell in the bulk, \maxb, and it has many extra
consequences,
as we have seen in the text. From the point of view of de Rham cohomology
the most immediate is that the dual current, $*j$, is not just coclosed
(current conservation, \cc) but coexact, and hence an element of the
subgroup
$K^{m-p-1}(\M,\B)$, \mbk,  of the relative de Rham cohomology group.
This is
the subgroup that
plays the role of kernel/image at this stage of the exact sequence of
cohomology. Thus the exact sequence is now brought into play
by means of the bulk Maxwell's equations.

When this current, $j$,  is determined by the geometrical picture
in terms of the $p$-brane world-volume, $w$, by \derham, the fact that
$\mu$ realises the Poincar\'e-Lefschetz isomorphism as
explained in the Appendix,
means that the world-volume $w$ must belong to a class of $K_{p+1}(\M)$,
\kb,
and hence be homologous to a cycle on the boundary, $\B$, of space-time.
This was one of our main results, obtained by a more roundabout,
though more self-contained, method,
when we were not taking the complete mathematical structure for granted.

This conclusion is contrary to what would have seemed intuitively likely,
that any configuration of brane world-volumes in space-time is possible.
The reason unsuitable configurations are forbidden is that they
provide topological obstructions to the integration of the
generalised Maxwell equations
for which they provide sources, as argued in the text. Notice that
in obtaining this selection rule
it was not necessary to take into account any equations of motion
for the brane degrees of freedom.

We saw that a another, related, consequence of Maxwell's inhomogeneous
equations in the bulk was the reduction of the count of linearly
independent
electric charges from the Betti number $b_{m-p-1}(\M,\B)$
to $s_{p+1}(\M)=b_{m-p-1}(\M,\B)-s_{m-p-1}(\M,\B)$,
corresponding to the number of linearly independent
homology classes permitted for the $p$-brane world-volume.

In section 4 we saw that $p$-brane electric fluxes are
classified by the boundary homology group, $H_{m-p-2}(\B;\Z)$, or,
more precisely by the free part of this abelian group obtained by dividing
out the torsion subgroup, namely a lattice of dimension given by the Betti
number $b_{m-p-2}(\B)$. The effect of the inhomogeneous bulk Maxwell
equations
is to equate to electric charges those electric fluxes on the sublattice of
dimension $s_{m-p-2}(\B)$,
corresponding to $K_{m-p-2}(\B;\Z)$. As a result
these electric fluxes are quantised, as integer multiples of $q$, but this
result does not apply the remaining $b_{m-p-2}(\B)-s_{m-p-2}(\B)$
electric fluxes. There is no reason for them to be quantised.

Through these boundary cycles there may also be  magnetic fluxes,
this time associated with $\tilde p$-branes, dual to the $p$-branes
(so $p+{\tilde p}+4=m$). As seen in section 6, these  vanish on the
afore-mentioned sublattice
of dimension $s_{m-p-2}(\B)$ whilst the remaining fluxes are quantised
as integer multiples of $2\pi\hbar/q$. Thus there is evidence of
states carrying just quantised electric charge and no magnetic charge,
and these must be the input $p$-brane states.
But the  quantised magnetic flux and non-quantised
electric flux through the $(H/K)_{m-p-2}(\B;\Z)$ cycles is
rather reminiscent of known solutions [Witten 1979] to
the Zwanziger-Schwinger quantisation condition
[Zwanziger 1968, Schwinger 1969] applying to particles
in four dimensional space-time and so provides evidence of mysterious
and intriguing dyonic objects
that are not situated on the space-time, $\M$, according to \maxa.
Maybe a better understanding of this phenomenon is important
in connection with electromagnetic duality.

At this stage, we should explain that one motivation for the
present work was to gain a better understanding of electromagnetic duality
[Montonen and Olive 1977]. It has been understood
that in a closed space-time of four dimensions, certain
partition functions exhibit a beautiful covariance under the action
of the modular group implementing electro-magnetic duality transformations
[Verlinde 1995, Witten 1995] and this is further enhanced when spin is
taken into account [Alvarez and Olive 2001]. It is also possible to
include Wilson loops [Zucchini 02]. But, in closed space-times there
are neither non-vanishing electric charges
nor electric fluxes, only magnetic fluxes. Yet in supersymmetric
gauge theories on flat space-time it is familiar
that electromagnetic duality transformations permute
electric and magnetic charges [Sen 1994].

So it might be important to consider space-times with boundary, as we have.
As just explained there is a beautiful topological classification
of conservation laws involving electric charge, electric flux and magnetic
flux but leaving no room for the classification of
magnetic charge. As a result we are left with a dilemma to be resolved
by future work.

There are many other questions left open for future work and many of them
concern undesirable simplifications that have been made relative to the
full
complication of superstring theory. We shall conclude by listing some of
these.
Some of these oversimplifications are routine practice in the subject
but should none-the-less be removed when possible.

1). Branes have been treated as geometrical objects, cycles in space-time,
and not assigned any sort of generalised quantum mechanical wave function
as ideally they should. In the absence of this  there is lacking the
concept
of intrinsic spin which is familiar for particle ($p=0$) wave-functions
on four dimensional space-time, and known to play a role in the
understanding
of electromagnetic duality [Alvarez and Olive 2001].

2) No account is taken of any internal brane structure, such as gauge
theories confined to the brane as sometimes required by supersymmetry.
If so presumably a $K$-theory classification of this internal structure
would be relevant, [Witten 2001].

3) Brane world-volumes have been treated as cycles. It might be more
reasonable
to allow them to have boundaries in the infinite past or future but we
do not know how to do this.

4) No equations of motion for $p$-brane degrees of freedom have been
considered. Partly this is because thse equations ought to 
involve the wave-functions, not yet formulated properly anyway when $p>0$.

5) No account is taken of any supersymmetry. This usually requires
a spectrum of values of $p$ and the fact that some branes may  possess
boundaries situated on other branes [Strominger 1996].
It would be interesting to know
how the charges and fluxes we have discussed could be related to the
tensor charges occurring in the supersymmetry algebra.

6) Branes have been treated as carrying only electric charge
but maybe an additional  magnetic charge should be allowed as an input
in \maxa.

7) No account of Chern-Simons type terms has been taken in the
generalised Maxwell equations. This could only occur when $p+2$ is even
and divides $m+1$, as for the familiar case of $p=2$ and $m=11$,
[Cremmer, Julia and Scherk 1978].

8) No special consideration has been made of the self-dual case when
$m$ equals twice  $p+2$ (so $p=\tilde p$).

\medskip
\section{Appendix}

The basic idea stems from the way the term in the action, \inter,
describing the geometrical coupling of the $p$-brane to the
gauge potential $A$, defines the dual electric current, $*j$,
via \intera\ to be proportional to a distribution valued
differential form, $\mu(w)$, depending on the world-volume, $w$.
Clearly this idea is motivated by physical considerations. A mathematical
version had earlier been proposed by de Rham [1955,1984].
So far the idea applies to absolute cycles and it has to be
extended to relative cycles and to chains both in the bulk
and on the boundary and this is now done.

If $C$ is a $q$-chain containing no sub $q$-chain
lying in the boundary, $\B$, its dual current, $\mu(C)$, is defined by
$$
\int_Cf=\int_{\M}f\wedge\mu(C)
$$ where $f$ is an arbitrary $q$-form.
On the other hand if $\gamma$ is a $q$-chain lying on the boundary, $\B$,
the dual surface current, $\nu(\gamma)$ is defined by
$$
\int_{\gamma}g=\int_{\B}g\wedge\nu(\gamma),
$$
where $g$ is an arbitary $q$-form on the boundary. Notice
that even though $C$ and $\gamma$ are chains of the same dimension, $q$,
$\mu(C)$ and  $\nu(\gamma)$ are forms of different degree,  $m-q$ and
$m-q-1$ respectively.

The boundary of $C$, $\partial C$, can be decomposed into two terms of
the type just described, each of dimension one less:
$$
\partial C=U+\alpha,
$$
so that $\mu(C)$, $\mu(U)$ and $\nu(\alpha)$ are all well-defined.

The integral $\int_{\partial C}h$ can be evaluated in two ways, first as
$\int_{\M}h\wedge\mu(U)+\int_{\B}h\atbound\wedge\nu(\alpha)$,
and secondly as $\int_{\M}dh\wedge\mu(C)$, using Stokes' theorem.
On integrating by parts this equals
$\int_{\B}h\atbound\wedge\mu(C)\atbound+(-1)^{d(C)}\int_{\M}h\wedge
d\mu(C)$,
where $d(C)$ is the dimension of $C$.
Equating the bulk and boundary terms separately yields the identities
$$
d\mu(C)=(-1)^{d(C)}\mu(U)\qquad\hbox{ and }
\qquad \mu(C)\atbound=\nu(\alpha).
$$
These are precisely what is needed to check the properties of the upwards
Poincar\'e-Lefschetz homomorphism from homology to de Rham
cohomology. This will be done by exploiting the different ways of
interpreting these relations and special cases of them.

$C$ is a relative cycle if $U$ vanishes. Then $d\mu(C)=0$ and so $\mu(U)$
is an absolute bulk de Rham cocycle as $\mu(C)\atbound$ need not vanish.

$C$ is an absolute bulk cycle if $U$ and $\alpha$ both vanish.
Then $d\mu(C)$ and $\mu(C)\atbound$ both vanish, implying that
$\mu(C)$ is a relative de Rham cocycle.

$U$ is a relative boundary and $\mu(U)$ is coexact and so
trivial in absolute bulk cohomology.

$U$ is an absolute boundary if $\alpha$ vanishes. Again $\mu(U)$ is coexact
but in addition $\mu(U)\atbound=\nu(\alpha)=0$ so that now $\mu(U)$
is trivial in relative cohomology.

These four observations are sufficient to show that $\mu$ maps absolute or
relative  homology classes into relative or absolute cohomology classes
respectively. By linearity these maps are homorphisms and it is easy to
see that their kernels include the torsion subgroups so more properly $\mu$
acts on the free parts, $F$, of the homology groups $H$ (obtained by
dividing out the torsion). So
$$
\mu:\qquad F_q(\M;\Z)\rightarrow F^{m-q}(\M,\B;\Z)\qquad\hbox{ and }
\qquad F_{q}(\M,\B;\Z)\rightarrow F^{m-q}(\M;\Z).
$$
The last step is to check that $\mu$ maps the
the appropriate kernel subgroups into each other.

$U$ is an absolute bulk cycle homologous to a boundary cycle, $-\alpha$ if
$\partial U$ vanishes and so in a class of $K_q(\M)$ by \km. But then
$\mu(U)=d[(-1)^{d(C)}\mu(C)]$ and $\mu(C)\atbound=\nu(\alpha)$ where
$d\nu(\alpha)=-(-1)^{d(C)}\nu(\partial\alpha)=0$. So, by \mbk, $\mu$
maps from a class of $K_q(\M)$ to a class of $K^{m-q}(\M,\B)$.

Finally if $U$ vanishes and $\alpha=\partial\gamma$,
then $\partial C=\partial\gamma$ meaning that $C$
is a relative cycle in a class of $K_q(\M,\B)$ by \kmb.
Hence $d\mu(C)$ vanishes and $\mu(C)\atbound=\nu(\alpha)=
\nu(\partial\gamma)=(-1)^{d(\gamma)}d\nu(\gamma)$. Thus, by \mk, $\mu$
maps from a class of $K_q(\M,\B)$ to a class of $K^{m-q}(\M)$.

By the work of this Appendix, the intersection number
$$
I(w,S)\equiv\int_S\,\mu(w)=\int_{\M}\mu(w)\wedge\mu(S).
$$
Furthermore if $w\in K_q(\M;\Z)$ and $S\in K_{m-q}(\M,\B;\Z)$
then $\mu(w)\in K^{m-q}(\M,\B;\Z)$ and $\mu(S)\in K^q(\M;\Z)$
so that, finally,  $I(w,S)$ vanishes by \van, as desired.
\bigskip

{\bf{Acknowledgements}}
\medskip
DI~Olive is belatedly grateful to G-C Wick for first introducing him to
homology theory, long ago, and to Tobias Ekholm and Victor Pidstrigach for
separately explaining important mathematical concepts to him.
He thanks  the Mittag-Leffler Institute (Djursholm), IFT (UNESP S\~ao
Paulo), the Yukawa Institute (University of Kyoto) and NORDITA
for hospitality whilst parts
of this work were accomplished.  M~Alvarez's research has been
supported by PPARC through the Advanced Fellowship PPA/A/S/1999/00486.
Support to both of us from the European String Network HPRN-CT-2000-122 is
also gratefully acknowledged.

\vfill\eject

\line{{\bf References}\hfil}
\medskip
\def\cmp{Commun Math Phys }\def\pl{Phys Lett } \def\np{Nucl Phys }

\parskip 3pt

\noindent M Alvarez and DI Olive,
\lq\lq The Dirac Quantisation Condition for Fluxes on Four-\hfil\break
Manifolds", {\tt hep-th 9906093}, \cmp {\bf 210} (2000) 13-28.

\noindent M Alvarez and DI Olive, \lq\lq Spin and abelian electromagnetic
duality on four-manifolds" {\tt hep-th/0003155},
\cmp {\bf 217} (2001) 331-356.

\noindent E Cremmer, B Julia and J Scherk,
``Supergravity Theory In 11 Dimensions,''
Phys.\ Lett.\ B {\bf 76} (1978) 409.

\noindent G de Rham; \lq\lq Vari\'et\'es Diff\'erentiables'', Hermann 1955;
\lq\lq Differentiable Manifolds'', Comprehensive Studies in Mathematics
{\bf 266} Springer 1984.

\noindent S Deser, A Gomberoff, M Henneaux and C Teitelboim,
\lq\lq Duality, self-duality, source and charge
quantisation in abelian $N$-form theories'',
Phys Lett {\bf B400} (1997) 80-86.

\noindent PAM Dirac, \lq\lq Quantised singularities in the
electromagnetic field'', Proc. Roy. Soc. {\bf A33} (1931) 60-72.

\noindent J Figueroa-O'Farrill and S Stanciu, \lq\lq D-brane charge,
flux quantisation and relative \hfil\break(co)homology'', {\tt
hep-th/0008032},
JHEP 0101:006 (2001).

\noindent H Flanders, \lq\lq Differential forms, with applications
to the physical sciences'', Academic 1963, Dover 1989.

\noindent M Henneaux and C Teitelboim,
\lq\lq $p$-form Electrodynamics'', Found  Phys {\bf 16} (1986) 593-717.

\noindent WVD Hodge, \lq\lq The theory and applications
of harmonic integrals'', CUP 1952.

\noindent M Kalb and P Ramond, \lq\lq Classical Direct Interstring
Action'',
Phys Rev {\bf D9} (1974) 2273-2284.

\noindent J Kalkkinen and K Stelle, \lq\lq Large gauge transformations
in M-theory'', {\tt hep-th/0212081}.

\noindent WS Massey, \lq\lq A basic course in algebraic topology'',
Graduate Texts in Mathematics {\bf 127} Springer 1991.

\noindent G Moore and E Witten, \lq\lq Self-duality, Ramond-Ramond
fields, and K-theory'', \hfil\break{\tt hep-th/9912279}, JHEP 0005:32
(2000).

\noindent A Schwarz, \lq\lq Topology  for Physicists'', Comprehensive
Studies in Mathematics {\bf 308},\hfil\break Springer 1994.

\noindent J ~S ~Schwinger,
``A Magnetic Model Of Matter,''
Science {\bf 165} (1969) 757.

\noindent A Sen; \lq\lq Dyon-monopole bound states,
self-dual harmonic forms on the multi-monopole moduli space, and
$SL(2,\Z)$ invariance in string theory", \pl\ {\bf B329} (1994), 217-221.

\noindent A Strominger, \lq\lq Open $p$-branes,
Phys Lett {\bf B383} (1996) 44-47 , {\tt hep-th/9512059}.

\noindent R Nepomechie, \lq\lq Magnetic monopoles from
 antisymmetric tensor gauge fields'', Phys Rev {\bf D31} (1985) 1921-1924.

\noindent C Teitelboim, \lq\lq Gauge invariance for extended objects'',
Phys Lett {\bf B167} (1986) 63-68.

\noindent E Verlinde; \lq\lq Global aspects of electric-magnetic duality",
\np\ {\bf B455} (1995), 211-228.

\noindent E~Witten,
``Dyons Of Charge $e\theta / 2 \pi$,''
Phys.\ Lett.\ B {\bf 86} (1979) 283-287.

\noindent E~Witten; \lq\lq On S-duality in abelian gauge theory",
{\it Selecta Math (NS)} {\bf 1} (1995), 383-410, {\tt hep-th/9505186}

\noindent E~Witten,
``Overview of K-theory applied to strings,''
Int.\ J.\ Mod.\ Phys.\ A {\bf 16} (2001) 693
{\tt hep-th/0007175}.

\noindent TT Wu and CN Yang, \lq\lq Concept of non-integrable phase factors
and global formulation of gauge fields'',
Phys. Rev. {\bf D12} (1975) 3845-3857.

\noindent R~Zucchini; \lq\lq Abelian duality and Wilson loops'',
{\tt hep-th/0210244}.

\noindent D~Zwanziger,
``Quantum Field Theory Of Particles With Both Electric And Magnetic
Charges,''
Phys.\ Rev.\  {\bf 176} (1968) 1489.

\end